\documentclass[aps,preprint,groupedaddress]{revtex4}
\usepackage{dcolumn}
\usepackage{amsmath}
\usepackage{graphicx}
\begin{document}

\title{The trivial Higgs boson: first evidences from LHC}

\author{P. Cea$^{1,2}$}
\email[]{Paolo.Cea@ba.infn.it}
\author{L. Cosmai$^{1}$}
\email[]{Leonardo.Cosmai@ba.infn.it}
\affiliation{$^1$INFN - Sezione di Bari, I-70126 Bari, Italy\\
$^2$Physics Department, Univ. of Bari, I-70126 Bari, Italy }

\begin{abstract}
We further elaborate on the triviality and spontaneous symmetry breaking scenario where the Higgs boson without self-interaction coexists with spontaneous symmetry breaking. 
The trivial Higgs boson  is rather heavy  with mass  $m_H = 754 \,  \pm \, 20 \, {\text{(stat)}}  \, \pm \, 20 \, {\text{(syst)}} \,  {\text{GeV}}$ and  total width  $\Gamma( H)  \, \simeq \, 320 \, {\text{GeV}}$. We briefly discuss the experimantal signatures of our trivial Higgs and compare with the recent results from ATLAS collaboration. We argue that experimental data
seem to support our scenario.
\end{abstract}

\pacs{14.80.Bn, 11.30.Qc,11.10.Hi}
\maketitle
A cornerstone of the Standard Model is the mechanism of spontaneous symmetry breaking that, as is well known, is mediated by the Higgs boson. Then, the discovery of the Higgs boson is the highest priority of the Large Hadron Collider (LHC). \\
\indent
Usually the spontaneous symmetry breaking in the Standard Model is implemented within the perturbation theory~\cite{Englert:1964,Higgs:1964,Guralnik:1964,Higgs:1966} which leads  to predict that the Higgs boson mass squared, $m^2_H$, is proportional to $\lambda_R \;  v^2_R$, where $v_R$ is the known weak scale (246~GeV) and $\lambda_R$ is the renormalized scalar self-coupling.  However, it is known   since long time that strictly local self-interacting four dimensional scalar field theories are trivial, namely $\lambda_R   \rightarrow 0$.
Quite recently~\cite{Cea:2009} we have enlightened  the scenario where the Higgs boson without self-interaction could coexists with spontaneous symmetry breaking. The  point is that,
due to the peculiar rescaling of  the Higgs condensate, the relation between $m_H$ and the physical $v_R$ is not the same as in perturbation theory. Indeed, according to this picture   one expects that  the ratio $m_H/v_R$ would  be a cutoff-independent constant. In other words,  one should have~\cite{Cea:2009}:
\begin{equation}
\label{1}
m_H \; = \;  \xi \;  v_R  
\end{equation}
where $\xi$ is  a constant. \\
\indent
It is noteworthy to point out that Eq.~(\ref{1}) can be checked by non-perturbative numerical simulations of self-interacting four dimensional scalar field theories on the lattice. Indeed, in previous studies~\cite{Cea:2009} we found numerical evidences in support of Eq.~(\ref{1}). Moreover, our numerical results showed that the extrapolation to the continuum limit leads to the quite simple result:
\begin{equation}
\label{2}
m_H \; \simeq \;  \pi  \;  v_R  \; 
\end{equation}
pointing to  a rather massive Higgs boson without self-interactions (triviality)~\cite{Cea:2009}:
\begin{equation}
\label{3}
m_H \; =  \; 754 \,  \pm \, 20 \, {\text{(stat)}}  \, \pm \, 20 \, {\text{(syst)}} \;  {\text{GeV}} \; .
\end{equation}

One could object that our lattice estimate of the Higgs mass  is not relevant for the physical Higgs boson. Indeed, the scalar theory relevant for the Standard Model is the O(4)-symmetric self-interacting theory. However, the Higgs mechanism eliminates three scalar fields leaving as physical Higgs field  the radial excitation whose dynamics  is described by  the one-component self-interacting scalar field theory. Therefore, we are confident that our determination of  the Higgs mass applies also to the Standard Model Higgs boson. \\

For Higgs mass in the range $700 - 800 \; {\text{GeV}}$ the main production mechanism at LHC is the gluon fusion $gg  \rightarrow H$. 
The gluon coupling to the Higgs boson in the Standard Model is mediated by triangular loops of top and bottom quarks. Since the Yukawa coupling of the Higgs particle to heavy quarks grows with quark mass, thus bilancing the decrease of the triangle amplitude, the effective gluon coupling approaches a non-zero value for large loop-quark masses. On the other hand, we already argued~\cite{Cea:2009} that the Higgs condensate rescaling suggests that,  if the fermions acquires a finite mass through the Yukawa couplings, then the coupling of the physical Higgs field to the fermions must vanishes or be  suppressed.   Fortunately, for large Higgs masses the vector-boson fusion mechanism becomes competitive to gluon fusion Higgs production. At $\sqrt{s} = 7 $ TeV we estimate:
\begin{equation}
\label{4}
\sigma(W^+ \, W^-   \rightarrow H) \; \simeq \; 0.03  -  0.05  \; {\text{pb}} \; \; , \; \;  700 \; {\text{GeV}} \; < m_H \; < \; 800   \;  {\text{GeV}} \; .
\end{equation}
The main difficulty in the experimental identification of a very heavy Standard Model Higgs ($m_H > 650 \; {\text{GeV}}$)  resides in the  large width which makes impossible to observe a mass peak.  However, in the triviality and spontaneous symmetry breaking scenario the Higgs self-coupling vanishes so that the decay width is mainly given by the  decays  into pairs of massive gauge bosons. Since the Higgs is trivial there are no loop corrections due to the Higgs self-coupling and we obtain for the Higgs total width:
\begin{equation}
\label{5}
\Gamma( H)  \; \simeq \; \Gamma( H \rightarrow W^+ \, W^-)  \; + \; \Gamma( H \rightarrow Z^0 \, Z^0) \; 
\end{equation}
where~\cite{Djouadi:2005gi}
\begin{equation}
\label{6}
 \Gamma( H \rightarrow W^+ \, W^-)   \; \simeq \;  \frac{G_F m_H^3}{8 \sqrt{2 \pi}} \, \sqrt{1 - 4 x_W} \, (1 \, - \, 4 x_W \, + 12 x_W^2) \; , \; x_W \; = \; \frac{m_W^2}{m_H^2} 
\end{equation}
\begin{equation}
\label{7}
 \Gamma( H \rightarrow Z^0 \, Z^0)   \; \simeq \;  \frac{G_F m_H^3}{16 \sqrt{2 \pi}} \, \sqrt{1 - 4 x_Z} \, (1 \, - \, 4 x_Z \, + 12 x_Z^2) \; , \; x_Z \; = \; \frac{m_Z^2}{m_H^2} \; . 
\end{equation}
Assuming $m_H \, \simeq \, 750 \; {\text{GeV}}$,  $m_W \, \simeq \, 80 \; {\text{GeV}}$ and $m_Z \, \simeq \, 91 \; {\text{GeV}}$, we obtain:
\begin{equation}
\label{8}
\Gamma( H)  \; \simeq \; 320 \; {\text{GeV}} \; . 
\end{equation}
\begin{figure}[t]
\includegraphics[width=0.9\textwidth,clip]{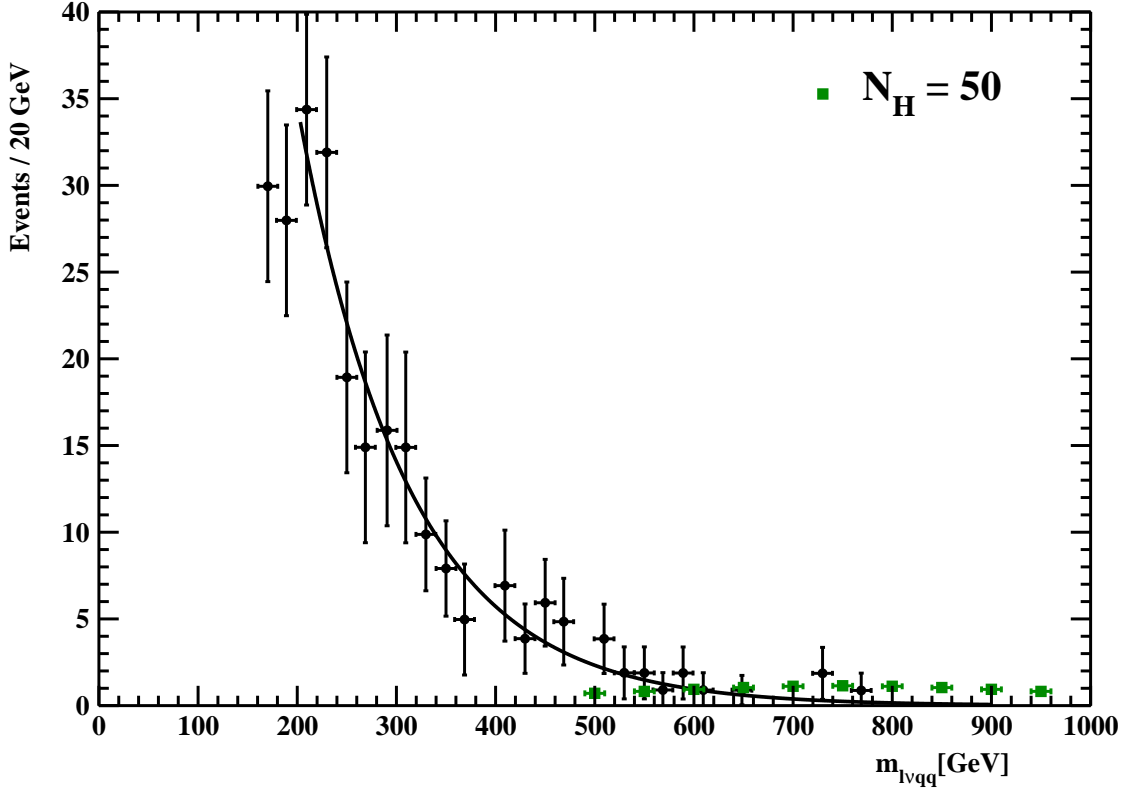}
\caption{\label{fig-1} Distribution of the invariant mass $m_{\ell \nu qq}$ for the  process $H \; \rightarrow WW \; \rightarrow \ell \nu qq$
corresponding to an integrated luminosity of 35 pb$^{-1}$. The data has been extracted from Fig. 4, panel b) of Ref~\cite{ATLAS:2011}.
The continuous line is a falling exponential function which models the background invariant mass spectrum. The squares are
the Higgs event distribution according to Eq.~(\ref{9}) with $N_H=50$  binned in energy intervals of 20 GeV.}
\end{figure}
Recently, the ATLAS collaboration~\cite{ATLAS:2011}  reported the experimental results for the search of the Standard Model Higgs boson at the Large Hadron Collider
running at $\sqrt{s} = 7$ TeV, based on a total integrated luminosity of about 40 pb$^{-1}$. In particular, in Fig.~\ref{fig-1} we display the distribution of the invariant mass for the
Higgs boson candidates corresponding to the process $H \; \rightarrow WW \; \rightarrow \ell \nu qq$. According to Ref.~\cite{ATLAS:2011}, the events were selected requiring exactly one lepton with $p_T \; > 30$ GeV. The missing transverse energy in the event were required to be $E^{miss}_T \; > \; 30$ GeV. The  invariant mass continuum  background is
parametrized as a  falling exponential function. To compare the invariant mass spectrum of our trivial Higgs with the experimental data,
we observe thast the energy distribution of the Higgs events is parametrized by the lorentzian distribution:
\begin{equation}
\label{9}
\frac{d \; n}{d \; E} \; = \; N_H \; \frac{1.15}{\pi} \; \frac{\Gamma( H)}{(E \; - \; m_H)^2 \; + \;  \Gamma( H)^2} \; , \; 
\Gamma( H)  \simeq  320 \; {\text{GeV}}  \;  \; , 
\end{equation}
where  $N_ H$ is the number of Higgs events.  In Fig.~\ref{fig-1} we compare  the lorentzian distribution of the invariant mass  binned in energy intervals of 20 GeV
assuming $m_H \, \simeq \, 750 \; {\text{GeV}}$ and  $N_ H = 50$, which would corresponds to an integrated luminosity of a few fb$^{-1}$.  For $m_{\ell \nu qq} \; > \; 700$ GeV the continuum background is
strongly suppressed, while the trivial Higgs event distribution is almost flat  up to 1000 GeV. It is remarkable that the experimental data do show an excess of three events in this
region. This compare quite well with our  estimate of  $\sigma(W^+ \, W^-   \rightarrow H)$, Eq.~(\ref{4}). In fact, tacking into account the uncertainties on the gluon-fusion production
mechanism and the decay branching ratio, we estimate about 1 - 2 Higgs events  for an integrated luminosity of 35 pb$^{-1}$.  Even though the very low statistics 
do not allow to draw definitive conclusions, we expect that by increasing the statistics the region $m_{\ell \nu qq} \; > \; 700$ GeV  will be almost uniformly
populated by Higgs events.\\
\indent
To conclude,  we proposed that strictly  local scalar fields are compatible with spontaneous symmetry breaking. In this case,  the Standard Model Higgs boson turns out to
be rather heavy.  We compared our proposal with the recent results from ATLAS collaboration and  argued that experimental data
seem to support our scenario. Moreover, we pointed out that our trivial Higgs boson scenario can be confirmed or rejected by simply increasing the statistics.
Since both the ATLAS and CMS collaborations have already collected an integrated luminosity of about 1 fb$^{-1}$, we expect that in the near future
the experimental data will corroborate our proposal.
\\
\indent
Finally, we would like to comment on the fact that our previous  paper was sent to a scientific journal for publication and
 was rejected by an anonymous referee with the following motivation:
\\
\noindent
{\em Therefore we can conclude that the analysis presented in the paper is simply 
not solid enough to corroborate the great claims about the Higgs mass in the SM.}
\\
\noindent
We decided to leave to LHC  the reply to the anonymous referee. Indeed, we feel that the time  is coming  to undertake  a profound
revision of the peer review process.

\end{document}